\documentstyle[a4,12pt]{article}

\newcommand{\beq}{\begin{equation}}
\newcommand{\eeq}{\end{equation}}
\def\beqa#1{\beq \begin{array}{#1}}
\newcommand{\eeqa}{\end{array} \eeq}

\def\g^a#1#2#3{\Gamma^{{#1}{#2}}_{#3}}
\def\g_a#1#2#3{\Gamma^{#1}_{{#2}{#3}}}

\def\slash#1{\rlap{\kern0.07em/}#1}			
\def\lslash{\rlap{\kern-0.03em/}l}			
\def\slashcap#1{\rlap{\kern0.25em/}#1}			
\def\hta{\rlap{\kern-0.04em/}h}				

\def\ma2#1#2#3#4{\left(\matrix{&{#1}&{#2}\cr&{#3}&{#4}}\right)}
\def\ma3#1#2#3#4#5#6#7#8#9{\left(\matrix{&{#1}&{#2}&{#3}\cr&{#4}&{#5}&{#6}
    \cr&{#7}&{#8}&{#9}}\right)}

\newcommand{\bbox}[1] { {\bf #1} }

\def\bbox#1{%
\relax\ifmmode
\mathchoice
{{\hbox{\boldmath$\displaystyle#1$}}}%
{{\hbox{\boldmath$\textstyle#1$}}}%
{{\hbox{\boldmath$\scriptstyle#1$}}}%
{{\hbox{\boldmath$\scriptscriptstyle#1$}}}%
\else
\mbox{#1}%
\fi
}
\begin{document}



\pagenumbering{arabic}

\begin{titlepage} \vspace{0.2in} \begin{flushright}
MITH-98/9 \\ \end{flushright} \vspace*{0.5cm}
\begin{center} {\LARGE \bf Glasses: a new view from QED coherence
\\} \vspace*{1.3cm}
{\bf M.~Buzzacchi, E.~Del Giudice, G.~Preparata}\\ \vspace*{0.5cm}
Dipartimento di Fisica dell'Universit\`a and I.N.F.N. - Sezione di Milano, 
Via Celoria 16, 20133 Milan, Italy\\ \vspace*{1.3cm}
{\bf   Abstract  \\ } \end{center} \indent
{\it Stuck in the marshes of the Kauzmann paradox, glasses have always been a 
puzzle for condensed matter theorists. We show that in the new picture of 
condensed matter, which takes into account the coherent interaction mechanisms
of QED, glasses are nothing but liquids, whose non coherent fraction is 
highly depleted, very close to zero near T$_g$, the temperature of 
glass formation. Using the recently developed QED theory of liquid water, we
are also able to give a successful account of the surprising finding of two
low-temperature water amorphs and of their phase-transition.}

\vfill \begin{flushleft} P.A.C.S. codes: 61.43.f, 64.70.p\vspace*{1cm} \\
\noindent{\rule[-.3cm]{5cm}{.02cm}} \\
\vspace*{0.2cm} \hspace*{0.5cm} 
E-mail addresses: Buzzacchi@mi.infn.it, Delgiudice@mi.infn.it, 
Preparata@mi.infn.it\end{flushleft} \end{titlepage}

\section{Introduction}

Physicists have been, and still are, greatly embarrassed when they are pressed
to define glasses, these elusive physical systems which look like solids, since
they keep, as crystals do, their volume and shape, but unlike crystals they
notably lack a microscopic space order. However, in spite of such apparent lack
of order, their entropy approaches that of crystals below a temperature  that
is usually rather high, a surprising fact that has been referred to as the
Kauzmann paradox \cite{Kauz}. For many substances, including water, the
Kauzmann temperature is higher than 100~K.

Quite apart from this "entropy puzzle", the very definition of the "glassy
state" gets entangled in some rather thorny theoretical problems. Whereas, in
fact, the transition from the glass to the solid is well behaved according to
the usual thermodynamical rules (the transition occurs on a sharply defined
line in a $p-T$ plane and is accompanied by the release of a well defined heat
of transition), the glass-liquid transition occurs as a gradual evolution in
phase space. Indeed, at a given pressure, the transition spans a temperature
interval, usually one or two tens of Kelvin wide. Sometimes, as in the case of
water, which we shall analyse in this paper, more than one glass phase exist
and the transition between them obeys the laws of orthodox thermodynamics,
being first-order.

Traditionally, glasses have been thought to belong to the realm of liquids,
since they admit a surface tension, as indicated by the increase of the length
of a fissure produced in a glass plate. Thus a glass could be considered as a
liquid possessing an infinite (actually an enormously high) viscosity. When the 
temperature decreases many liquids, including water, exhibit a rapid increase
in viscosity, associated with the divergence of all relaxation times
\cite{Ang}. As a result, it is a matter of convention to define which is the
threshold of viscosity, beyond which the liquid becomes a glass or,
alternatively, how long the experimentalist should wait before declaring the
formation of the glass.

The slowing down of many thermodynamical transformations that has been observed
when a piece of matter approaches the "glassy state" is at the center of what
Anderson \cite{And} has termed a "change of paradigm", in Kuhn's acception
\cite{Kuhn}. According to this view, the enormous increase in viscosity just
blocks the evolution of the macroscopic system in its phase-space, so that the
system, so to say, gets stuck in one of its microscopic configurations,
becoming unable to "explore" the full ensemble of its allowed configurations. 
In other words, the system loses its "ergodicity", thus making the concept of
entropy void.

In order to obtain an understanding of such "localization in phase-space", as
Anderson has christened this concept, the potential energy surfaces are
pictured as steeply discontinuous landscapes. When kinetic energy, i.e.
temperature, is low enough, the system finds it more and more difficult to
migrate from one to the next, thus causing an increase of viscosity. According
to the new paradigm, localization in phase-space should explain also the
Kauzmann paradox, since in that space the system "knows" its own small "hole"
and nothing else and, as a result, its "effective" entropy is indeed very
small.

We believe, however, that this interpretation of the thermodynamics of glasses
runs into some difficulties when confronted with the ease with which the low
density amorph (LDA, a glass of H$_2$O) makes a transition towards the recently
discovered \cite{Mish} high density one (HDA). Since this transition occurs 
between two glasses below the Kauzmann temperature (where their entropy is very
small), it must be dynamically driven, i.e. the change $\Delta G$ of the Gibbs
potential must be mainly due to a change of enthalpy $\Delta H$ , induced by
applying a large external pressure. Thus we must accept that for glasses there
are two types of transformation: one associated with diffusive processes that
become slower and slower with the decrease of temperature; the other induced by
changes in non-thermal variables (such as pressure) that occur within
"reasonable" time spans, in which significant density changes are observed
($\rho_{LDA}\simeq 0.92$~g/cm$^3$, $\rho_{HDA}\simeq 1.3$~g/cm$^3$) in spite of
the small change in $G$ ($\Delta G \simeq 2.5$~kJ/mole, corresponding to
26~meV per molecule), indicating that during the transition the system did not
have to climb very steep slopes in phase space. This dual nature of glass
transformations brings to mind an old fashioned model of glasses, the so-called
"extended Glarum model" \cite{Gla},\cite{Phil}, which pictures the glass as a
basic non-entropic "continuum" containing a number of "defects" capable to
diffuse among a number of "sites". With the decrease of temperature, the
"defects" and the "sites" become scarcer and scarcer, thus lengthening the
diffusion times. In such model the transport properties depend on the
"defects", while the "continuum" contributes essentially to the structure which
is assumed to be quite heavy, thus rather insensitive to molecular agitation.
However, this "defect and hole" theory has been criticized (see, for instance,
Ref.\cite{And}) on the ground that it is hard to reconcile the assumed
"continuum" with the molecular structure that, according to the generally
accepted picture of condensed matter, forms the glass.

In recent times a new approach to condensed matter has been developed, based on
Quantum Electrodynamics (QED) \cite{Prep}, that allows us to give a fresh look
to all the above problems. In this theory an ensemble of molecules or atoms,
beyond a critical density and below a critical temperature, become unstable
against rearranging itself in a new ground state, where the molecules
become a coherent matter field that oscillates, in tune with the e.m. field,
between two selected molecular levels. In this way the molecules restructure
themselves since their configuration becomes a superposition of the two
molecular levels. The minimum space-domain where such coherent oscillations
take place, the Coherence Domain (CD), has the size of the wavelength of the
electromagnetic mode resonantly coupled to the transition between the two
molecular configurations. At room temperature the thermal fluctuations extract
from the "coherent ground state" a number of molecules that can't any longer
follow the coherent oscillation: they appear and behave as a dense gas
occupying the interstices among the CD's, forming a non-coherent fraction,
whose size decreases with temperature. We can thus see that the non-coherent
fraction plays the same role as the "defects" of the phenomenological Glarum
model.

Within such framework, we have analysed the dynamics and thermodynamics of
water in detail \cite{Ara}, obtaining good agreement between theory and
experiment. In particular we have found that the non-coherent fraction (see 
Fig.6 of Ref.[9] and Fig.1 of this paper) practically vanishes at a temperature 
as high as 135~K: one can see that a decrease from $1\%$ to $0.1\%$ is obtained 
in the interval 150$\sim$120~K, thus reproducing a very peculiar feature of the 
"glassy transition".

A careful inspection of Fig.1 renders it reasonable to identify the "glass"
with the liquid where the coherent fraction has become just "very close" to
one. Of course, in this configuration viscosity becomes enormous, for in a
coherent system it is impossible to move one molecule without affecting all the
others, that are phase-related to it. However, contrary to the conventional
vision, high viscosity (and in particular the high observed deviations from 
Arrhenius activation) is the \underline{consequence} and not the
\underline{cause} of the formation of glass, due to the fact that 
the allowed phase-space
shrinks to the coherent ground-state, losing the entropy of the incoherent
phase. One may say that in the new view the glass is not just a "piece of
ill-condensed matter" as Anderson puts it, but on the contrary it is 
matter in a well
condensed state, where all molecules oscillate in unison and are not 
bound to be
spatially ordered. The hindered translational degrees of freedom of the glass 
make its
entropy negligible, thus explaining the Kauzmann paradox in a natural way.
Even though throughout this paper we shall illustrate our general theory of 
glasses in the only case we have studied in sufficient depth, i.e. liquid
water, the basic ideas we have just discussed must also be applicable to other
glass-forming systems, whose basic molecular structure may be quite different
from that of H$_2$O.
 
This much in the way of Introduction. The rest of the paper is organized as
follows: in Sect.2 we summarize the main results of Ref.[9]; Sect.3 discusses
within the coherent QED framework the principal features of a glass, while Sect.4
contains a derivation of the thermodynamics of the transition between the two
amorphs of water.

\section{QED structure of water}

In the framework laid down in Refs.[8,9] we have worked out a theory of liquid
water, whose main points we shall now summarize.

The electronic spectrum of water is very rich (see Fig.2), making it rather
difficult to select the level $|B\rangle$, that is going to be the partner of
the ground state $|0\rangle$ in the coherent oscillations induced by the
quantized electromagnetic field. As explained at length in Chap.3 of Ref.[8],
the main ingredients of a coherent process are:
\begin{itemize}
\begin{enumerate}
\item the direct transitions $|0\rangle \leftrightarrow |B\rangle$, whose
amplitude is governed by the coupling constant:
\beq
g_{B}=\left(\frac{2\pi}{3}\right)^{1/2}\frac{e}{m_{e}^{1/2}}
\left(\frac{N}{V}\right)^{1/2}\frac{f_{0B}^{1/2}}{\omega_{B}},
\eeq
where in the natural units system ($\hbar=c=k_{B}=1$) $m_e$ is the electron
mass, $N/V$ the number density of the system, $\omega_B$ the energy difference
$E_{B}-E_{0}$, and $f_{0B}$ the oscillator strength of the transition  
$|0\rangle \leftrightarrow |B\rangle$;

\item the "photon mass" term $\mu(\omega)$, that arises from the virtual
transitions from the ground state to all other intermediate states $|n\rangle$,
induced by the photon modes of frequency $\omega$, whose value is 
\beq
\mu_{B}(\omega)=-\frac{1}{2} \frac{e^2}{m_{e}}\left(\frac{N}{V}\right)\frac{1}
{\omega_{B}^2}\sum_{n\neq B}f_{0n}\frac{\omega^2}{\omega_{n}^{2}-\omega^{2}}.
\eeq
\end{enumerate}
\end{itemize}
Defining $\mu_{B}(\omega_{B})=\mu_{B}$, one finds [8] that, when
\beq
g_{B}^{2}\geq g_{B,crit}^{2}=\frac{8}{27}+\frac{2}{3}\mu_{B}+
\left(\frac{4}{9}+\frac{2}{3}\mu_{B}\right)^{3/2},
\eeq
the ensemble of molecules becomes dynamically unstable: the molecules begin to
oscillate between  $|0\rangle$ and $|B\rangle$ in a spatial 
region whose minimum size
is the wavelength of the resonant electromagnetic mode:
\beq
\lambda_{B}=\frac{2\pi}{\omega_{B}},
\eeq
which will be called the "Coherence Domain" (CD) of the system. The oscillation 
 $|0\rangle \leftrightarrow |B\rangle$, in tune with the corresponding e.m.
mode, will dynamically evolve (see Ref.[8]) to a "renormalized" frequency:
\beq
\omega_{r}< \omega_{B},
\eeq
that guarantees that the e.m. field \underline{stays where the molecules are},
and is not radiated away, thus decreasing the energy of the ground state: an
impossible undertaking!

Eq.(3) determines the minimum critical density for which the transition from
the "perturbative" ground state to the "coherent" ground state occurs. Thus
each level $|B\rangle$ has its own "critical density" and the competition among
the levels will obviously be gained by the level, let's call it  $|1\rangle$,
that has the lowest critical density, since at this density the ensemble of
molecules will start oscillating collectively between  $|0\rangle$ and 
$|1\rangle$, preventing all other levels from participating in the "e.m.
dance". As it will be shown in Sect.4, the influence of macroscopic variables,
such as pressure, upon the terms appearing in Eq.(3) may lead the ground state 
$|0\rangle$ to change partner in the coherent oscillation, giving rise in such
way to phase-transitions, that considerably enrich the phase diagram. It is
interesting to note that at zero pressure Eq.(3) selects among the different
levels of the H$_2$O molecule the excited state at 12.06~eV, whose critical
density is 0.88~g/cm$^3$.

The theory of Refs.[8,9] shows that for $g\simeq g_{crit}$, the e.m. potential
(averaged over different directions) $A$ obeys the equation 
(differentiating with respect to the adimensional time $\tau=\omega t$)
\beq
\frac{i}{2}\frac{d^{3}A}{d\tau^{3}}+\frac{d^{2}A}{d\tau^{2}}
+i\mu\frac{dA}{d\tau}+g^{2}A=0,
\eeq
that, for very small $g^{2}$, admits for $\mu<-0.5$ (and thus
$\rho>\rho_{crit}$) runaway solutions, namely $A$ grows exponentially, reaching
in times of the order of 10$^{-14}$~sec a limiting value corresponding to a
physical configuration where:
\begin{itemize}
\begin{enumerate}
\item the electron clouds of the 
water molecules are described by the coherent state:
\beq
|coh.\rangle=\cos \gamma |0\rangle + \sin \gamma |1\rangle,
\eeq
with $\cos^{2}\gamma=0.873$. Should one wish to produce a (incoherent)
superposition of this type in a thermal way, he would need an oven kept at 
$7\cdot10^{4}$~K and would be, of course, compelled to accept heavy
contamination from many other levels;
\item
the oscillation of the matter field is in tune with a corresponding oscillation
of the e.m. field. According to (5), the common frequency of oscillation is
lower (indeed, much lower) than the original value $\omega=12.06$~eV at the
start of the runaway process. While the wavelength $\lambda$ remains unchanged,
the "renormalized" frequency attains the remarkably reduced value:
\beq
\omega_{r}=0.26~eV.
\eeq
As a consequence this dynamically generated e.m. field cannot be radiated,
since the mass of its "photon", instead of vanishing, has an imaginary value:
\beq
m^2=\omega_{r}^{2}-\vec{k}^{2}\ll \omega^{2}-\vec{k}^{2}=0.
\eeq
We note that the renormalization of the frequency (Eq.8) is absolutely crucial
for the consistency of the full physical picture, since by Eq.(9) the e.m.
field is prevented from leaving the system and lower further its energy. Thus
this dynamical configuration does represent the state of minimum energy.

\item
The overall energy gain that such configuration achieves can be understood as
follows. The positive energy that is required to produce the coherent e.m.
field and to excite 12.7$\%$ of the molecules to the level at 12.06~eV is more
than compensated by the negative energy of the interaction between the e.m.
field and the current generated by the oscillating molecules; the sum of these
three terms becomes negative when the density exceeds a critical threshold,
becoming more and more negative with increasing density. As a consequence, the
molecules crowd as close as they are allowed by the highly
repulsive short-range forces originating from the molecular hard cores. Since the excited
state $|12.06~eV\rangle$ is spatially quite more extended than the ground state
$|0\rangle$, the intermolecular distance is larger than would be predicted from
the standard molecular size. In Ref.\cite{Delg} this problem has been
thoroughly discussed. The energy gained in the process of condensation is
released out and accounts for the first-order character of the vapour-liquid
transition. The analysis of Ref.[9] shows that the energy gain $\Delta E$ per
molecule at the center of a CD is
\beq
\Delta E=\omega_{r}=0.26~eV.
\eeq
Due to the space variation of the e.m. vector potential $A$, $\Delta E$
exhibits the modulation as a function of the distance $r$ from the center of
the CD shown in Fig.3. Moreover, the equilibrium condition between the 
molecular and the e.m. fields demands that the molecules are assembled 
in the central part of the CD up 
to a radius which is 3/4 of the total radius, given by Eq.(4).

\item
Given an ensemble of CD's, energy minimization requires that the fields of 
the different domains be in tune. This phase-matching allows the fields in 
the interstices to interfere to yield energy minimization. In such configuration 
the molecules at the boundary of each CD are acted upon not only 
by the e.m. field 
of their own CD but also by the tails of the fields of neighbouring CD's. 
The profile of the total energy is depicted in Fig.4. In this way the coherence 
of a single domain is propagated throughout condensed matter on a macroscopic 
scale.

\item
At T$\neq$0 the coherent ground state is subject to the thermal aggression of 
the environment, absorbing energy and momentum from the collisions with external 
particles or from external radiation. These energy-momentum transfers excite single 
molecules out of the coherent state to one of the single-particle states 
described by an appropriate excitation curve. In Fig.5 we report the curve 
used in Ref.[9]. Thus for each T we may compute the non-coherent fraction 
$F_{nc}(T)$ of molecules extracted from the coherent ground state , the coherent 
fraction being obviously given by $F_{c}(T)=1-F_{nc}(T)$. In Fig.1 we plot 
$F_{c}(T)$, as derived from the excitation spectrum assumed in Ref.[9]. 
The non-coherent fraction comprises the molecules extracted from the regions 
where $|\Delta E|$ is smaller. An inspection of Fig.4 shows these regions to 
be located at the periphery of the CD's which, as a result, shrink when T 
increases. Thus we may write for the "effective" radius of the CD as a function 
of T:
\beq
R_{CD}(T)=R_{CD}(0)[F_{c}(T)]^{1/3},
\eeq
where Eq.(4) informs us that $R_{CD}(0)=\frac{3\pi}{4\omega_{0}}$. Please 
note, however, that the two fractions are not sharply divided, but their 
boundaries are somewhat blurred, so that the transition from the coherent to 
the non-coherent fraction is rather smooth.

The requirement of phase-matching among CD's for energy minimization keeps the 
CD's tightly packed, in such a way that the non-coherent fraction remains 
trapped in the interstices of an array of CD's, that form a kind of cage 
preventing the non-coherent "gaseous" phase to leave the system as a free gas. 
Of course, the non-coherent fraction is not an ideal gas, but rather a Van der 
Waals' gas, since significant short range attractive forces act between the 
molecules. When T increases, the non-coherent fraction increases its pressure, 
which appears as the vapour tension of the liquid. At the boiling point the 
pressure of the non-coherent fraction succeeds in breaking the "cages" of the  
CD's and in coming to the open. Thermodynamical equilibrium then requires that 
a new "non-coherent phase" be established at the expense of the coherent one; 
all the energy supplied to the system from the outside gets spent to accomplish 
just 
this. The new non-coherent phase also finds its way out of the liquid, and the 
process keeps on going until the complete vaporization of the liquid is 
achieved.

To end this Section, we would like to recall that, apart from the
successful analysis of the dynamics and thermodynamics of water carried out in
Ref.\cite{Ara}, QED finds another remarkable corroboration from the striking
phenomena of Single Bubble Sonoluminescence \cite{Putterman}, which we have
shown to arise from the electrodynamic nature of the interactions among the
molecules of H$_2$O in Ref.\cite{Buzz}.
\end{enumerate}
\end{itemize}
\section{H$_2$O at low temperature: supercooled water and glass}

In this Section we shall discuss some peculiar properties of liquid water at low 
temperature, as predicted by the QED theory outlined in the previous Section, 
according to which most molecules now belong to the coherent fluid.

As a consequence of the two-fluid picture of water characteristic of our 
approach,
all physical variables are given by the weighted combination of the "coherent-
phase" value, which is independent of temperature, and the value in the 
"non-coherent phase", typical of a dense gas. For instance, in the case of an 
extensive variable $X$, such as the free energy or the entropy, one has:
\beq
X(T)=X_{c}F_{c}(T)+X_{nc}(T)F_{nc}(T),
\eeq
where $X_{c}$ is independent of temperature, since the internal temperature 
of the coherent phase is zero. The thermodynamic potentials, like the specific 
heat, derivatives with respect to temperature of extensive quantities, can thus 
be written: 
\beq
\frac{dX}{dT}(T)=[X_{c}-X_{nc}(T)]\frac{dF_{c}}{dT}+F_{nc}(T)\frac{dX_{nc}}{dT}.
\eeq

At low temperature $F_{nc}(T)$ can be neglected, so that the thermodynamic 
potentials are proportional to $\frac{dF_{c}}{dT}$, whose temperature 
behaviour can be seen in Fig.1b. It is interesting to compare this behaviour with 
that of the typical specific heat of a glass, shown in Fig.6. Both show a 
characteristic $\lambda$-shape: the curve increases steeply in the region of the 
supercooled liquid and drops to zero below the "glassy transition". In our approach 
this peculiar shape is a consequence of the T-dependence of the filling by the
non-coherent fraction of the interstices of the CD's.

In order to understand the dynamics of the "glassy transition", let us analyse 
the behaviour of the shear viscosity $\eta$ at low temperature, where the 
non-coherent fraction $F_{nc}(T)$ has become quite small. The main point is 
that the coherent fluid behaves like an infinitely viscous liquid, since below 
the gap threshold it is impossible to accelerate a single molecule without 
collectively accelerating the whole CD. Thus the observed viscosity of the 
liquid should, according to our theory, be ascribed to the non-coherent fluid 
that lingers in the interstices between CD's: it is just the flow of the 
non-coherent fluid, which acts as a kind of lubricant, which allows the array 
of CD's to follow the stream of the liquid. When $F_{c}(T)\rightarrow 1$ 
$(F_{nc}(T)\rightarrow 0)$ the "lubricant" disappears, viscosity becomes 
very large and the liquid vitrifies. It is clear, therefore, that this 
process is not sharp in temperature, like the usual thermodynamic phase 
transitions, but it actually starts when the non-coherent fraction becomes 
small enough to make viscosity exceed a fixed large value. In order to have 
a rough idea of the phenomenon, we consider the crude model where the CD's are 
rigid clusters of water molecules floating in the non-coherent fluid, streaming 
in a channel of width $R$ and where the size of the average interstice between 
CD's is $\delta$. From the usual definition of shear viscosity $\eta$, one 
gets:
\beq
\frac{\eta_{tot}(T)}{R}=\frac{\eta_{nc}(T)}{\delta},
\eeq
hence
\beq
\eta_{tot}(T)=\frac{\eta_{nc}(T)}{\delta /R}\propto \frac{\eta_{nc}(T)}{F_{nc}(T)},
\eeq
implying that when $F_{nc}\rightarrow 0$ the viscosity $\eta_{tot}(T)$ will 
diverge. Inspection of Fig.1a shows that our model for $F_{nc}(T)=1-F_{c}(T)$ 
approaches the vanishing limiting value for $T\simeq 135$~K. In Fig.7 the interval 
$100~K<T<200~K$ of Fig.1a has been magnified: one can see that at 150~K 
$F_{nc}\simeq 1.6\cdot 10^{-2}$, whereas at 135~K $F_{nc}\simeq 10^{-3}$, 
showing that by decreasing the temperature by approximately 15~K $\eta$ 
increases by an order of magnitude. Thus the non-sharp nature of the glassy 
transition is just the consequence of the non-sharp disapperance of the non-
coherent fluid.

The above discussion allows us also to understand the large increase of 
viscosity observed \cite{Pesc} when a shear stress is exerted upon a layer 
of water whose thickness is smaller than a threshold value, which at room 
temperature is about $500~\AA$. We note that at this temperature Eq.(11) yields 
for the CD diameter just such value; thus when the width of the channel is 
about the CD diameter no room is left for the "lubricant" and $\eta$ is bound 
to rise very sharply.

All other transport variables, from the self-diffusion coefficients to the 
thermal relaxation times, which critically depend on $F_{nc}(T)$, are expected 
to parallel the behaviour of $\eta$, thus producing the kind of universal 
behaviour that is typical of all supercooled liquids just above vitrification. 

The problem of the divergence of viscosity at low temperature has been 
recently  
analysed for a Lennard-Jones liquid \cite{Par} in the conventional framework of 
the free energy landscapes discussed in the Introduction. It is interesting to
observe that for the viscosity and the relaxation time these authors derive 
a behaviour very close to the 
Arrhenius one, which is almost universally inadequate to describe 
the observed transport properties near the glassy 
transition, the Kauzmann paradox and the $\lambda$-shape of heat capacity. 
It is quite reasonable that the dynamics discussed in Ref.[14] is at work
within the non-coherent fluid, producing the divergence of $\eta_{nc}$ in
Eq.(15). However, it seems equally reasonable that the non-Arrhenius behaviour
of the viscosity is the necessary consequence of the fast (in temperature) 
depletion near $T_{g}$ 
of the non-coherent phase and, thus, can only be understood in a
conceptual framework which recognizes the role of coherence.

\section{The tale of two glasses: LDA and HDA}

As mentioned in the Introduction, a challenge to the usual understanding of 
liquid has been posed by the discovery of a high density variety of amorphous 
solid water \cite{Mish}. The existence of a liquid-liquid phase transition, 
which gives rise to a kind of "liquid polymorphism", has been suggested in 
Refs.\cite{Mish2},\cite{Bell}. Hence, the generally accepted conceptual 
framework is hard put to explain how two different intermolecular organizations 
may arise from the same basic molecule, capable of well defined electric 
polarizations, sources of equally well defined fields of force. Moreover, 
since the phase transition occurs at very low temperature, where diffusive 
processes are very slow, the transformation must be dynamically driven, for the 
application of an appropriate pressure transforms a mesoscopic metamolecular 
complex into a different one in a short time in a non-diffusive way.

We now show that such stringent requirements, strongly suggested by the empirical 
evidence, are naturally accounted for in the QED approach we have followed in 
this paper. In Ref.[9] it was found that the excited level at 12.06~eV, 
"winner" of the competition for the partnership with the ground state in the 
coherent, collective oscillation, is actually closely trailed in the race by a 
level at 11.5~eV, whose critical density is at 1~g/cm$^3$. This level, 
whose electronic configuration is $4p$, is less extended than the "winner", 
which is $5d$. Should the level at 11.5~eV have won the race, the resulting 
liquid 
would clearly have been denser than the actual liquid water. In Table~1 we  
summarize the main features of the two levels.
\begin{center}
\begin{tabular}{|c|c|c|} \hline
 & Level 1 & Level 2 \\ \hline 
$\omega$/eV & 12.06 & 11.5 \\ \hline
electronic configuration & $5d$ & $4s$ \\ \hline
$\rho$ (g/cm$^3$) & 0.88 & 1.0 \\ \hline 
\end{tabular}
\end{center}
\begin{center}
{\bf {\it Table 1.}}
\end{center}
At low temperature, where due to the smallness of the non-coherent fraction 
the entropic contribution is negligible, the free energy per molecule $f$ is 
the sum of three contributions:
\beq
f=\frac{F}{N}\simeq \frac{E}{N}=-\delta_{c}-\delta_{SR,attr}+\delta_{SR,rep}
\eeq
which we are now going to briefly discuss:
\begin{itemize}
\begin{enumerate}
\item -$\delta_{c}$ is the gap produced by the coherent oscillation between 
the ground state and the chosen level. From the discussion in Ref.[8]
\beq
\delta_{c}=\omega \epsilon,
\eeq
where $\epsilon$ satisfies the quartic equation:
\beq
2\epsilon^{4}+\epsilon^{3}-3\epsilon^{2}(1+\mu_{r})-
\epsilon(2g^{2}+1+2\mu_{r})+(1+2\mu_{r})^{2}=0.
\eeq
Since $\delta_{c}$ is expected from thermodynamics to be in the range of  
tenths of eV, whereas $\omega$ exceeds 10~eV, $\epsilon$ must be quite small, 
thus allowing us to drop in Eq.(18) $\epsilon$-powers higher than~1, yielding:
\beq
\epsilon \simeq \frac{(1+2\mu_{r})^2}{1+2g^{2}+2\mu_{r}}.
\eeq
For the level $|1\rangle$ one can show that this approximation differs from 
the exact solution by less than 5$\%$. Also, in the region of $n=N/V\simeq 
1 \frac{g}{m_{H_{2}O}cm^{3}}$ the gap $\delta_{c}$ has a very mild dependence 
on $n$, so that we shall assume $\delta(n)\simeq$~const.

\item
$\delta_{SR,attr}$ denotes the attractive effect of the short-range (Van der 
Waals) interaction, which prevails at distances not too close to the molecular 
hard core. From the energy functional
\beq
E_{SR}[n]=\frac{1}{2}\int d^{3}\vec{x}~d^{3}\vec{y}~
\psi^{\dagger}(\vec{x})\psi(\vec{x})V_{SR}(\vec{x}-\vec{y})
\psi^{\dagger}(\vec{y})\psi(\vec{y}),
\eeq
where $|\psi|\simeq n^{1/2}$ and the potential is free from singularities, 
we get:
\beq
\delta_{SR,attr}=\alpha n^{2}.
\eeq

\item
$\delta_{SR,rep}$ arises from the interaction between the hard cores, which repel 
each other due to the combined effects of the Coulomb repulsion and of the Pauli 
principle. Using the repulsive part of a typical Lennard-Jones interaction:
\beq
V_{SR,rep}(r)=\epsilon \left(\frac{r_{0}}{r}\right)^{12},
\eeq
and carrying out the analysis reported in Ref.\cite{Neut}, 
we obtain the functional dependence
\beq
\delta_{SR,rep}=\beta n^{7/3}.
\eeq
\end{enumerate}
\end{itemize}
Thus Eq.(10) takes the form:
\beq
f=-\delta_{c}-\alpha n^{2}+\beta n^{7/3}.
\eeq
From the equation of state:
\beq
p=-\frac{\partial F}{\partial V}=
-\frac{\partial (F/N)}{\partial (V/N)}=n^{2}\frac{\partial f}{\partial n},
\eeq
we get:
\beq
p=n^{3}\left[-2\alpha +\frac{7}{3}\beta n^{1/3}\right],
\eeq
that for $p=0$ yields:
\beq
\alpha=\frac{7}{6}\beta n_{0}^{1/3},
\eeq
$n_{0}$ being thus the density at zero pressure. By combining (25) and (17), we 
obtain:
\beq
f=-\delta_{c}-\alpha n^{2}\left[1-\frac{6}{7}\left(\frac{n}{n_0}\right)^{1/3}\right],
\eeq
which at $p=0$ reduces to
\beq
f_{0}=-\delta_{c}-\frac{\alpha}{7}n_{0}^{2}.
\eeq
In addition we have
\beq
p=2\alpha n^{3}\left[\left(\frac{n}{n_0}\right)^{1/3}-1\right]. 
\eeq
Let us now analyse the problem of the two water glasses in the above framework. 
We have (See Ref.[9]) $\delta_{c}^{LDA}=0.26~eV$ and 
$\delta_{SR}^{HDA}=0.24~eV$ at $p=0$. Thus: 
\beq
f_{LDA}=-\delta_{c}^{LDA}-\delta_{SR}^{LDA}
\left[7-6\left(\frac{n}{n_{0}^{LDA}}\right)^{1/3}\right],
\eeq
\beq
f_{HDA}=-\delta_{c}^{HDA}-\delta_{SR}^{LDA}
\left(\frac{n_{0}^{HDA}}{n_{0}^{LDA}}\right)^{2}
\left[7-6\left(\frac{n}{n_{0}^{HDA}}\right)^{1/3}\right],
\eeq
\beq
p_{LDA}=14\delta_{SR}^{LDA}n
\left(\frac{n}{n_{0}^{LDA}}\right)^{2}
\left[\left(\frac{n}{n_{0}^{LDA}}\right)^{1/3}-1\right].
\eeq
Note that in the above equations the coefficient $\alpha$ of the Van der Waals 
attraction, being independent of density has been assumed as universal. Eq.(33) 
gives us in a straightforward manner the pressure of the transition 
LDA$\rightarrow$
HDA when we equate the density $n$ to the threshold value prescribed by eq.(3) for 
the level $|2\rangle$ at 11.5~eV (see also Table 1). By putting 
$\rho_{0}^{LDA}=0.94$~g/cm$^3$ and $\rho_{crit}^{(2)}=1$~g/cm$^3$, eq.(33) gives 
us:
\beq
p_{LDA\rightarrow HDA}\simeq 3.4~kbar,
\eeq
in excellent agreement with th value $p_{LDA\rightarrow HDA}=3.2$~kbar 
measured by Bellissent-Funel \cite{Bell}.

From the thermodynamic measurements of Ref.\cite{Bell} we are able to derive 
the parameters of the coherent process involving level $|2\rangle$, that 
gives rise to the HDA. By equating the Gibbs potentials per molecule
\beq
g=\frac{G}{N}=f+\frac{pV}{N}=f+\frac{p}{n}
\eeq
of the two glasses, we get:
\beq
\delta_{c}^{HDA}=
\delta_{c}^{LDA}-\delta_{SR}^{LDA}
\left[\left(\frac{n_{0}^{HDA}}{n_{0}^{LDA}}\right)^{2}-1\right]+
\frac{p}{n_{0}^{LDA}}\left(\frac{n_{0}^{LDA}}{n_{0}^{HDA}}-1\right).
\eeq
Inserting in this equation $p_{HDA\rightarrow LDA}=0.5$~kbar, measured in 
Ref.[16] and $\rho_{0}^{HDA}=1.17$~g/cm$^3$, we get:
\beq
\delta_{c}^{HDA}=0.13~eV=\frac{1}{2}\delta_{c}^{LDA},
\eeq
in good agreement with the value that can be obtained from Eq.(17).

Thus we find that in the HDA the electrodynamical gap $\delta_{c}^{HDA}$ is  
weaker than in the LDA, but at high pressures one can obtain higher densities 
with less energy expense, since the size of the molecular levels involved in 
the coherent oscillations selected by QED is smaller.

\section{Conclusions}

In this paper we have attempted to show that the new picture of condensed matter 
afforded by a full utilization of the general equations of QED, and in 
particular of their coherent solutions, provides us with an approach to the 
physics of the glassy state that seems both simple and powerful. In a nutshell, 
according to our theory, a glass is nothing else than a liquid, i.e. a molecular 
system whose coherent oscillations involve the valence electrons, which has 
(almost) completely lost its non-coherent fraction, thus becoming enormously 
viscous. Instead of ill-condensed matter, as Anderson sees it, a glass appears 
to us a perfectly condensed liquid. We have shown that such view is supported 
by the peculiar T-dependence of the function $F_{nc}(T)$, the non-coherent 
fraction, that we have determined for water (but whose qualitative structure 
should be quite general), which reproduces the typical shapes of the specific 
heat near the "glassy transition" and the basic undefiniteness of $T_g$, the 
glass transition temperature. The extreme viscosity of the glass is thus seen as 
the manifestation of the phase coherence of the matter field, which suppresses 
all local motions, which would result in large fluctuations of the phase itself.

In the last Section we have been able to give a successful theory of the non-
diffusive transition between the LDA and the HDA of water, simply in terms 
of the pressure induced change of the coherent oscillations of the water 
molecule. We have identified a different excited level  at 11.5~eV that, being 
smaller in size than the one at 12.06~eV, 
involved in the coherent oscillations 
of liquid water, leads to a denser glass. The quantitative agreement with the 
experimental observations of Ref.[16] is, we believe, a further corroboration 
of the power, simplicity and correctness of this approach, which we hope to 
extend soon to other interesting types of glass.

\end{document}